\documentclass[a4paper,11pt]{article}
\pdfoutput=1
\usepackage{jcappub}
\usepackage{amsmath}
\usepackage{amssymb}
\usepackage{CJK}
\usepackage{graphicx}% Include figure files
\usepackage{dcolumn}% Align table columns on decimal point
\usepackage{bm}% bold math
\usepackage[dvipsnames,table]{xcolor}
\hypersetup{urlcolor=BlueViolet,
	    citecolor=Plum,
	    linkcolor=PineGreen}
\newcommand{\xbar}{\overline{x}}

\begin{document}
% \preprint{CPPC-2020-05}
% \preprint{KCL-PH-TH/2020-43}
\subheader{\hfill CPPC-2020-05}

\title{Eliminating the LIGO bounds on primordial black hole dark matter}
\author[a]{C\'eline B\oe hm,}
\author[a]{Archil Kobakhidze,}
\author[a]{Ciaran A. J. O'Hare,}
\author[a]{Zachary S. C. Picker,}
\author[b]{Mairi Sakellariadou}
\affiliation[a]{Sydney Consortium for Particle Physics and Cosmology, \\
 School of Physics, The University of Sydney, NSW 2006, Australia }

\affiliation[b]{Theoretical Particle Physics and Cosmology Group, Physics Department,
King’s College London, University of London, Strand, London WC2R 2LS, UK}
\emailAdd{celine.boehm@sydney.edu.au}
\emailAdd{archil.kobakhidze@sydney.edu.au}
\emailAdd{ciaran.ohare@sydney.edu.au}
\emailAdd{zachary.picker@sydney.edu.au}
\emailAdd{mairi.sakellariadou@kcl.ac.uk}

\abstract{Primordial black holes (PBHs) in the mass range $(30$--$100)~M_{\odot}$ are interesting candidates for dark matter but are tightly constrained by the LIGO merger rate. In deriving these constraints, PBHs were treated as \emph{constant} Schwarzschild masses. A careful analysis of cosmological black holes however leads to a time-dependent effective mass. This implies stricter conditions for binary formation, so that the binaries formed merge well before LIGO’s observations. The observed binaries are those coalescing within galactic halos, at a rate consistent with LIGO data. This reopens the possibility of LIGO mass PBH dark matter.

}

\maketitle

\section{Introduction}
Primordial black holes (PBHs)~\cite{pbh} are one of the oldest and arguably least speculative candidates for the elusive dark matter~\cite{Hawking:1971ei, Carr:1974nx,Chapline:1975ojl}. They attracted a renewed interest recently~\cite{Bird:2016dcv, Sasaki:2016jop, Clesse:2016vqa} after the discovery by the LIGO/Virgo~Collaboration~\cite{Abbott:2016blz} of the mergers of black hole binaries in the mass range $\sim (10$--$50)~M_{\odot}$. Remarkably, the merger rate of virialized PBH binaries in this mass window is consistent with the rate inferred from the LIGO/Virgo observations~\cite{Bird:2016dcv}.
Moreover, black holes with $(30$--$100)~M_{\odot}$ masses sit in a narrow window between much more stringent stellar microlensing ~\cite{Oguri18,Allsman:2000kg,Tisserand:2006zx} and cosmic microwave background~\cite{Ricotti:2007au, Ali-Haimoud:2016mbv} constraints on the PBH dark matter fraction. 
While there do exist several bounds on the fraction of PBH dark matter that overlap this mass range --- e.g.~from dynamical effects on stellar systems~\cite{Monroy-Rodriguez:2014ula,Brandt:2016aco, Koushiappas:2017chw,Lu:2020bmd,Stegmann:2019wyz,Green:2016xgy,Li:2016utv}, extragalactic lensing~\cite{Mediavilla2009,Mediavilla:2017bok,Munoz:2016tmg,Liao:2020wae,Zumalacarregui:2017qqd}, or from PBH accretion at early~\cite{Hektor:2018qqw,Poulin:2017bwe,Horowitz:2016lib} and late~\cite{Manshanden:2018tze,Gaggero:2016dpq, Inoue:2017csr} times --- these often come with loopholes~\cite{Brandt:2016aco, Ali-Haimoud:2016mbv,Bosch-Ramon:2020pcz} and heavy astrophysical uncertainties~\cite{Green:2017qoa,Quinn2009,Zhu:2017plg}. Taking just these constraints, it would be plausible that what LIGO/Virgo has observed is indeed the fraction of the dark matter comprised of PBHs.

Here we focus on the constraints discussed in refs.~\cite{Sasaki:2016jop, Ali-Haimoud:2017rtz,Kocsis:2017yty} which seem to disprove much more convincingly the claim that LIGO/Virgo has observed PBH binaries merging. While the merger rate of \emph{virialized} PBHs is consistent with the LIGO/Virgo observations~\cite{Bird:2016dcv}, it was shown~\cite{Nakamura:1997sm, Ioka:1998nz} that PBH binaries would be produced at a much higher rate in the early universe. Recent estimates of the binary production rate~\cite{Sasaki:2016jop, Ali-Haimoud:2017rtz} imply that LIGO/Virgo should observe many more mergers than they do if PBHs constitute a large fraction of the dark matter. These calculations suggest that the PBH dark matter fraction must be $f_{\mathrm{PBH}}\lesssim 10^{-3}$ in the LIGO-mass range.\footnote{Clustering of PBHs may loosen these constraints \cite{Vaskonen:2019jpv,Raidal:2018bbj}. Several recent papers~\cite{Jedamzik:2020ypm,Jedamzik:2020omx,Atal:2020igj} have also claimed the constraints from the merger rate can be fully relaxed if PBH clustering takes place. This conclusion is complementary to ours but requires additional assumptions. Here we make no assumption about clustering, which is a more conservative option for our argument.} Up to date summaries of constraints on PBH dark matter can be found in, for example, refs.~\cite{Carr:2020gox,Green:2020jor}.

In this work, we readdress these constraints by questioning the choice of metric that most adequately describes PBHs in the early universe. In particular we focus on the period of time between their formation and their ``decoupling": the moment when the PBH binary's gravity overwhelms the outward pull of cosmological expansion and they evolve via local gravitational dynamics only. The choice of metric that most consistently describes black holes in expanding spacetime turns out to be far from trivial~\cite{McVittie:1933zz,Einstein:1945id,Faraoni:2018xwo}. Here, we argue that the Thakurta metric~\cite{Thakurta} is perhaps singled out as the most justifiable option, with drastic phenomenological consequences for PBHs. The key new feature of the Thakurta metric that distinguishes our treatment from the majority of prior PBH literature, is that the black holes it describes possess a time-dependent quasi-local effective mass. This feature changes nothing fundamental about the PBHs themselves, but rather how they appear to gravitate when in an expanding background. Therefore, this forces us to reconsider the requirements for a PBH binary to decouple, which, as we will show, means that the LIGO binary abundance constraints are completely evaded.

A summary of our result is the following. Under the assumption that the Thakurta metric models early-universe PBHs, a relatively straightforward set of arguments shows that there are much more stringent conditions for two PBHs to form a decoupled binary. As a result, very few binaries are formed in the early universe, and they only form when the lone PBHs are much closer together than one would conclude when adopting the crude treatment of black hole decoupling used previously. Once formed, the binaries are governed by local dynamics only (as per the definition of decoupling), and the time of coalescence is given by a simple formula that is a function of their initial separation. However, since Thakurta PBH binaries must be much closer together to form, the coalescence times are necessarily much shorter. This means that the rare binaries that do form will coalesce rapidly, well before the mergers observed by LIGO/Virgo. In fact, even if we consider binaries forming as late as redshift $z \sim 1$, they still merge well before the earliest observed LIGO/Virgo merger. This means that the only binaries that are present around the redshifts that LIGO/Virgo observe mergers happening, must be those that have found each other in virialized galactic halos. As we discussed above, the merger rate of this population of binaries is tantalizingly consistent with the observed merger rate~\cite{Bird:2016dcv}.

To begin, in section~\ref{sec:metrics} we define our treatment of PBHs following the Thakurta metric, discuss the features of this choice, and compare it with its alternatives. Then in section~\ref{sec:decoupling} we derive two conditions that PBH binaries must satisfy to decouple from cosmological expansion. This then leads to a calculation of PBH binary formation in section~\ref{sec:pbhformation} before we finally numerically evaluate those constraints for PBH binaries forming at a given redshift in section~\ref{sec:conditions}. We present concluding remarks in section~\ref{sec:conc}.

%Armed with a more realistic description of cosmological black holes, it is pertinent to consider recalculating the dynamics of their binaries. We detail these calculations in the following sections, but present our final result in figure~\ref{fig:Coalescencetime}. We show the resulting coalescence time, $\tau_b$, that a binary has if it decouples from the Hubble flow at some redshift $z_{\rm dec}$. This result stems from a new dynamical requirement for a PBH \emph{binary} to decouple from the Hubble flow---rather than just a single black hole as in the case of the standard decoupling condition. Our new requirement states that the orbital decay of the PBH binary due to gravitational wave emission must exceed the outward pull of cosmological expansion. This is a dynamical requirement that has not been considered previously and only appears as a consequence of the aforementioned feature of Thakurta black holes: namely, that they should have growing Misner-Sharp masses. From the moment of decoupling we use a completely standard description of the binary using only local dynamics: the coalescence time, $\tau_b$, is only dependent on the initial separation of the PBHs. The key difference in this work is that our new dynamical decoupling condition results in PBHs binaries that can only decouple when they are much closer together than calculated under the standard decoupling condition. This means that regardless of the decoupling redshift, our PBHs must coalesce much faster than before.

\section{Metrics to describe cosmological PBHs}\label{sec:metrics}
The LIGO binary abundance constraints rely on calculations showing that PBHs form binaries very efficiently in the early universe. This calculation usually proceeds by representing Schwarzschild black holes as point masses. However, in the early universe, the Hubble horizon is still small and so one cannot ignore the fact that these PBHs are embedded in a Friedmann-Lema\^{i}tre-Robertson-Walker (FLRW), as opposed to Minkowski, background. Thus the Schwarzschild solution cannot be an adequate description of these PBHs---we must use a cosmologically-embedded black hole. Only once the Hubble horizon is sufficiently large can we can safely assume that this cosmologically-embedded solution interpolates cleanly to the Schwarzschild solution. In fact, the usual binary abundance calculations already discuss such a decoupling time: it occurs when the mass-energy density of the PBH is larger than the background density. This condition should be kept in mind when we evaluate the various options for metrics to describe cosmological PBHs. 

\subsection{The Thakurta metric}
Regardless of the process of PBH formation, a suitable metric will need to somehow describe a central inhomogeneity which asymptotically approaches FLRW spacetime. In attempting to find a solution that does this, we run into the more fundamental issue of how cosmological expansion influences local gravitating systems in general; a problem with a long history dating back to the works of McVittie~\cite{McVittie:1933zz} and Einstein and Straus~\cite{Einstein:1945id}. For a recent review see, e.g.~ref.~\cite{Faraoni:2018xwo} and references therein. 

With several options available, one particular example, the Thakurta metric~\cite{Faraoni:2018xwo}, stands out since it is the late-time attractor solution of the entire class of generalized McVittie geometries~\cite{Faraoni:2007es}. These geometries are sourced by an imperfect fluid, with a radial heat flow. Even more generically, the Thakurta metric appears to be the general-relativistic limit of a class of exact solutions of Brans-Dicke gravity with a cosmological fluid~\cite{clifton10.1111/j.1365-2966.2005.08831.x}. The Thakurta metric is refreshingly simple; its line element can be written in terms of the Schwarzschild line element and cosmological scale factor $a(t)$ as,
\begin{align}
    \mathrm{d}s^2 = a^2(t) \, \mathrm{d}s^2_{\rm schw} \, .
\end{align}
In terms of the ``physical'' radial coordinate $R=a(t)r$ and angular coordinates $\theta, \phi$, this metric can be written as,
% \begin{align}\label{Thakurta}
% \mathrm{d}s^2&=f(R)\left(1-\frac{H^2R^2}{f^2(R)}\right)\mathrm{d}t^2 + \frac{2HR}{f(R)}\mathrm{d}t \, \mathrm{d}R \nonumber \\
% &-\frac{\mathrm{d}R^2}{f(R)}-R^2\left(\mathrm{d}\theta^2+\sin^2\theta \,\mathrm{d}\phi^2\right)\,,
% \end{align} 
\begin{equation}\label{Thakurta}
\mathrm{d}s^2=f(R)\left(1-\frac{H^2R^2}{f^2(R)}\right)\mathrm{d}t^2 + \frac{2HR}{f(R)}\mathrm{d}t \, \mathrm{d}R-\frac{\mathrm{d}R^2}{f(R)}-R^2\left(\mathrm{d}\theta^2+\sin^2\theta \,\mathrm{d}\phi^2\right)\,,
\end{equation} 
where $f(R)=1-2Gm a(t)/R=1-2Gm/r$ and $m$ is the physical mass of the PBHs today. The Hubble expansion rate reads $H=\dot{a}/a$, with $\dot{a}$ standing for the derivative of the scale factor. In accordance with physical intuition, at scales much smaller than the cosmological horizon, $(2Gma\lesssim)~ R \ll 1/H$, eq.(\ref{Thakurta}) approaches the static Schwarzschild spacetime, while at scales larger than the Schwarzschild radius $2Gma \ll R~(\lesssim1/H)$, the metric approaches the FLRW metric.

\subsection{Comparison to alternatives}
In contrast to the Thakurta metric, the original McVittie geometry~\cite{McVittie:1933zz} does not have any radial energy flow. For realistic FLRW backgrounds, this leads to a \emph{spacelike} naked singularity instead of the usual black hole horizon. In the perfect fluid which sources these geometries, a spacelike singularity corresponds to a divergent pressure, unless the universe asymptotically approaches a state of de Sitter expansion~\cite{Kaloper:2010ec}. PBHs, however, decouple well before our universe reaches this de Sitter phase of dark energy domination, hence the applicability of the original McVittie metric to the problem at hand is questionable.

Another option could be the Einstein-Straus~\cite{Einstein:1945id} metric in which a Schwarzschild black hole is simply stitched on to an expanding FLRW metric at some radius. This metric has no immediately obvious physical problems, however it is wholly inadequate for describing PBH decoupling. As we mentioned above, one can estimate the epoch of decoupling by finding the time at which the black hole mass density exceeds the density of the cosmological background. However stitching the boundary of the Einstein-Straus black hole to its surroundings actually enforces the mass density to be equal to background density. So it is impossible to describe the decoupling of Einstein-Straus black holes at all.

Finally, the Lema\^{i}tre-Tolman-Bondi class of metrics~\cite{Lemaitre:1933gd,Tolman:1934za,Bondi:1947fta}, which were originally proposed to model an expanding spherically-symmetric dust-filled universe, also contains a branch of collapsing solutions that could describe an emerging black hole. However, solutions that interpolate between the collapse at small scales and expansion at large scales are plagued by shell-crossing singularities~\cite{Joshi:2014gea}.      

Other analytic solutions than the ones mentioned above also tend to have physical problems such as negative energy densities, or singularities. We refer to ref.~\cite{Faraoni:2018xwo} for a more complete discussion. Ultimately, the Thakurta metric eq.(\ref{Thakurta}) appears to be  the simplest and least problematic of all the possible solutions. From here, we will simply choose this metric, and proceed by calculating the consequences of this choice. We emphasize though that apart from this choice, we adopt an entirely conventional treatment of both cosmologically embedded metrics and the binary abundance. For clarity, we first isolate the key feature of the Thakurta metric that has the most profound impact on the phenomenology of PBHs: the Misner-Sharp mass.

\subsection{The Misner-Sharp mass}
In a cosmological setting, the notion of the black hole mass requires a more careful definition. The most common definition is the Arnowitt-Deser-Misner (ADM) mass \cite{Arnowitt:1959ah}, which is defined at spatial infinity. However, PBHs are produced from the gravitational collapse of inhomogeneities at scales comparable to the Hubble radius: a scenario where there is no spatial infinity at which to define the ADM mass. For spherically symmetric cosmological black holes, instead of the ADM mass, it is more appropriate to define a quasi-local Misner-Sharp mass~\cite{Misner:1964je}, which is a measure of active mass in a given volume of spacetime. For the Thakurta metric eq.(\ref{Thakurta}), this mass can be directly calculated, or read-off from the 00-component of the metric:
\begin{equation}\label{MS}
m_{\mathrm{MS}}=ma(t)+\frac{H^2R^3}{2Gf(R)}~. 
\end{equation}
It is worth noting that, although we used a specific metric to compute the Misner-Sharp mass, it is known to be invariant under coordinate re-parametrization~\cite{cahill} and thus is a physically meaningful quantity. 

After PBHs are produced during the radiation dominated era, $H\sim 1/a^2$ and $R\sim a$. So while the first term in eq.(\ref{MS}) increases $\propto a(t)$, the second term decreases $\propto 1/a(t)$. Therefore we can ignore the latter and simply describe the PBH mass relevant for describing decoupling as being equal to $ma(t)$. We emphasize that this notion of mass makes no assumption about the constituents of the cosmological fluid, and its increase is \textit{not} due to any accretion of matter. The Misner-Sharp mass does not imply that any fundamental property of the black hole is changing. Rather, it is an effective, local mass: a feature of a geometry in which nearby test masses follow geodesics affected both by the central inhomogenity \textit{and} the exterior FLRW background. We expand on this idea in the next section.

%the concept of mass in General Relativity is not universal but rather depends on the background geometry. The Arnowitt-Deser-Misner (ADM) mass \cite{Arnowitt:1959ah} for vacuum black holes used in refs.~\cite{Nakamura:1997sm, Sasaki:2016jop, Ali-Haimoud:2017rtz} is strictly applicable only if spacetimes are asymptotically flat. For example, the ADM mass approximation is good when the typical size or time scale under consideration is much smaller than the scale at which the curvature of the background geometry becomes significant, like the case of stellar black holes. However, PBHs  are produced from the gravitational collapse of inhomogeneities at scales comparable to the Hubble radius: a scenario in which the ADM mass approximation is not applicable. Consequently, binary dynamics should be affected by the cosmological expansion, leading to a dramatically different merger rate than has been reported previously~\cite{Nakamura:1997sm, Sasaki:2016jop, Ali-Haimoud:2017rtz}. 

\section{Decoupling Conditions}\label{sec:decoupling}
\subsection{Static decoupling}
First, we should review the PBH decoupling condition that the \textit{usual} binary abundance calculations arrive at. This condition amounts to requiring that the inward force towards the black hole exceeds the outward pull of cosmological expansion. Mathematically this can be derived from the radial geodesic equation for a test particle in the FLRW universe augmented by the Newtonian force due to a central black hole:
\begin{equation}
    \ddot R=-\frac{Gm}{R^2}+\frac{\ddot a}{a}R \, .
\end{equation}
Where in the usual calculation, $m$ is the ADM mass of the black hole. By demanding that the attractive Newtonian force dominates over cosmological drag, one can straightforwardly show the standard decoupling condition is equivalent to the requirement that the local mass density around the black hole exceeds the energy density of the cosmic fluid. Despite being written in the context of an expanding universe, this static decoupling condition is a Newtonian calculation. 

Now, let us inspect the radial geodesic motion of a test mass for the Thakurta metric instead. In the relevant approximation $Gma \ll R \ll 1/H$ the geodesic equation is subtly different (see also ref.~\cite{Perez:2019cxw}):
  \begin{equation}\label{geo}
\ddot R=-\frac{Gma}{R^2}+\frac{\ddot a}{a}R \, . 
\end{equation}
The first term on the right hand side of this equation is the standard attractive Newtonian force exerted by the cosmological black hole of active mass $ma=m/(1+z)$. The second term in eq.(\ref{geo}) is entirely due to the cosmological expansion and has a simple heuristic interpretation: it is the cosmological drag force (per unit mass) acting on a test particle with no peculiar motion.  Ensuring that the Newtonian force dominates over the cosmological force, the ``static'' decoupling condition requires,
  \begin{align}\label{dec1}
\frac{ma}{V}\gg & \frac{3}{4\pi G}\left|\frac{\ddot a}{a}\right| \nonumber \\
&=\rho_{\rm cr}\left|-\Omega_m(1+z)^3-2\Omega_r(1+z)^4+2\Omega_{\Lambda}\right|~,
\end{align}
where, $V=(4\pi/3)R^3$ and $\Omega_{m,r,\Lambda} = \rho_{m,r,\Lambda}/\rho_{\rm cr}$ are respectively the cosmological densities of matter, radiation and a cosmological constant relative to the critical density, $\rho_{\rm cr}$. We assume $\Omega_m\approx 0.307$, $\Omega_{\Lambda}\approx 0.691$ and $\Omega_r\approx 5.4\times 10^{-5}$~\cite{Ade:2015xua}.

The static decoupling condition, eq.(\ref{dec1}), is very similar to the one used previously~\cite{Nakamura:1997sm, Ioka:1998nz,Sasaki:2016jop, Ali-Haimoud:2017rtz}, but differs in one important aspect: it involves the approximate time-dependent Misner-Sharp mass, $ma=m/(1+z)$, instead of the static ADM mass $m$. This has important ramifications for the decoupling condition of binaries. Since the Misner-Sharp mass grows with the cosmological expansion, it allows for binary decoupling at much later times compared to previous calculations, even allowing for decoupling later than matter-radiation equality. 

\subsection{Dynamical decoupling}
The modified static PBH decoupling condition already implies a weakening of the LIGO-mass PBH dark matter bounds. However, on closer inspection it turns out that the Misner-Sharp mass assumption requires another stronger decoupling condition to be written down, further weakening the bounds.

The argument is as follows: on the one hand, a binary system loses its energy through the emission of gravitational waves, resulting in orbital decay and eventually a merger.  On the other hand, cosmological expansion tends to pull black holes apart. Hence for a decoupled binary to merge, orbital decay due to the emission of gravitational waves must dominate over the pull of the expansion of the universe. Using the active Misner-Sharp mass, $m a(t)$, a system of two cosmological black holes separated by a radial distance $R$ carries the total energy $E=-GM\mu a^2/(2R)$ (in the Newtonian circular orbit approximation; for elliptical orbits $R$ refers to the semi-major axis), where $M=m_1+m_2$ and $\mu = m_1m_2/M$, are respectively the present-day total and reduced masses of the binary system. Hence, the change in radial separation of black holes is given by $\dot R/R=-\dot E/E  +2H$. Requiring that the binary be coalescing gives us a new ``dynamical'' condition for decoupling:
  \begin{align}\label{dec2}
\dot E/E & > 2H \nonumber \\
&=2H_0\sqrt{\Omega_M(1+z)^3+\Omega_r(1+z)^4+\Omega_{\Lambda}}~,
\end{align}
where $H_0=\sqrt{8\pi G\rho_{\rm cr}/3}$. At leading order, $\dot E$ is given by the suitably modified quadrupole formulae, which will be discussed in the next section.  

This condition is unique to PBHs with the growing quasi-local masses given by the Thakurta solution (in previous PBH calculations, this condition is met automatically).  In fact, the need for such a condition to describe binary formation can also be arrived at numerically. Simulated binary PBHs will not coalesce unless this condition is met. In practice this means that two PBHs must be much closer together to form a binary than one would expect following previous calculations. Therefore, this dynamical decoupling condition is the one that will have the most dramatic consequences for the early-universe PBH merger rate.

\section{PBH binary formation in the early universe}\label{sec:pbhformation} 
Now that we have a potentially important condition that must be met for a black hole binary to decouple and merge, we must consider what this implies for a population of PBHs forming in the early universe.

We will assume a random (as opposed to clustered) initial distribution and a monochromatic mass function at $m$. Both of these are probably unrealistic simplifications, but we will see that they are actually on the conservative side for the argument we present here. A population of PBHs constituting a fraction $\rho_{\mathrm{PBH}}=f_{\rm PBH}\rho_{\rm DM}$ of the cosmological dark matter density, would have an average separation today (at $t_0$) of~\footnote{Note that this quantity in the literature is usually defined at the epoch of matter-radiation equality ($z=z_{\rm eq}\sim 3000$), $\xbar=\xbar_0/(1+z_{\rm eq})$, since PBH binary decoupling in the matter dominated era was considered unfeasible.},
\begin{equation}
\xbar_0 = \left(\frac{m}{f_{\rm PBH}\rho_{\rm cr}\Omega_{\rm DM}}\right)^{1/3}\approx \frac{1.2~\mathrm{kpc}}{f_{\rm PBH}^{1/3}}\left(\frac{m}{30M_{\odot}}\right)^{1/3} \,.
\end{equation}
We then consider a pair of neighboring PBHs separated by a distance $x=x_0/(1+z)$ at redshift $z$, where $x_0$ is the separation at $t_0$. The third nearest PBH to the pair would help them to form an elliptical binary system through tidal effects, with semi-major axis $\mathfrak{a}$ and semi-minor axis $\mathfrak{b}$, given respectively by:
\begin{equation}\label{axis}
\mathfrak{a}=\alpha \frac{x_0}{1+z}~,~~\mathfrak{b}=\beta \left(\frac{x_0}{y_0}\right)^{3}\mathfrak{a}~,
\end{equation}
where $y=y_0/(1+z)$ is the distance to the third nearest black hole.  Numerical 3-body calculations (for a static mass) indicate that $\alpha\approx 0.4$ and $\beta\approx 0.8$ \cite{Ioka:1998nz} are suitable corrections for the semi-analytic approximation here.

The binary is considered to be formed when the system decouples from the Hubble flow at some $z=z_{\mathrm{dec}}$ and evolves according to local dynamics only, independent of the cosmological expansion. This is when both conditions, eqs.(\ref{dec1}) and (\ref{dec2}), are satisfied. After decoupling, the binary system with the semi-major axis $\mathfrak{a}_{\mathrm{dec}}=\mathfrak{a}(z_{\mathrm{dec}})$ and eccentricity $e_{\mathrm{dec}}=e(z_{\mathrm{dec}})$ coalesce with a lifetime given by the usual result,
\begin{equation}\label{life}
\tau_{\rm b}=\frac{3}{85}
\frac{\mathfrak{a}^4_{\mathrm{dec}}(1-e_{\mathrm{dec}}^2)^{7/2}}{r_{\rm s}^3}~,
\end{equation}
after which the PBHs merge~\cite{Peters:1964zz}. Here for convenience we define the binary Schwarzschild scale $r_{\rm s}=(G^3M^2\mu)^{1/3}$. Note that the cosmological evolution of binaries we describe here is only relevant before galaxy formation, so certainly not later than $z\sim 1$. At later times PBHs are expected to be bound within galactic halos, so that binaries start to form according to the scenario described in ref.~\cite{Bird:2016dcv}. The cosmic time elapsed from $z\sim1$ until a typical LIGO/Virgo binary merger at $z\sim 0.1$ is $\sim 6.6~\textrm{Gyr}$. Hence, the binaries formed over the course of cosmological evolution must live long enough to contribute to the merger rate observed by LIGO/Virgo, i.e.~$\tau_{\rm b}\gtrsim6.6~\textrm{Gyr}$.      

\section{Conditions for binary decoupling}\label{sec:conditions}

\begin{figure}[t]
    \centering
    \includegraphics[width=0.8\textwidth]{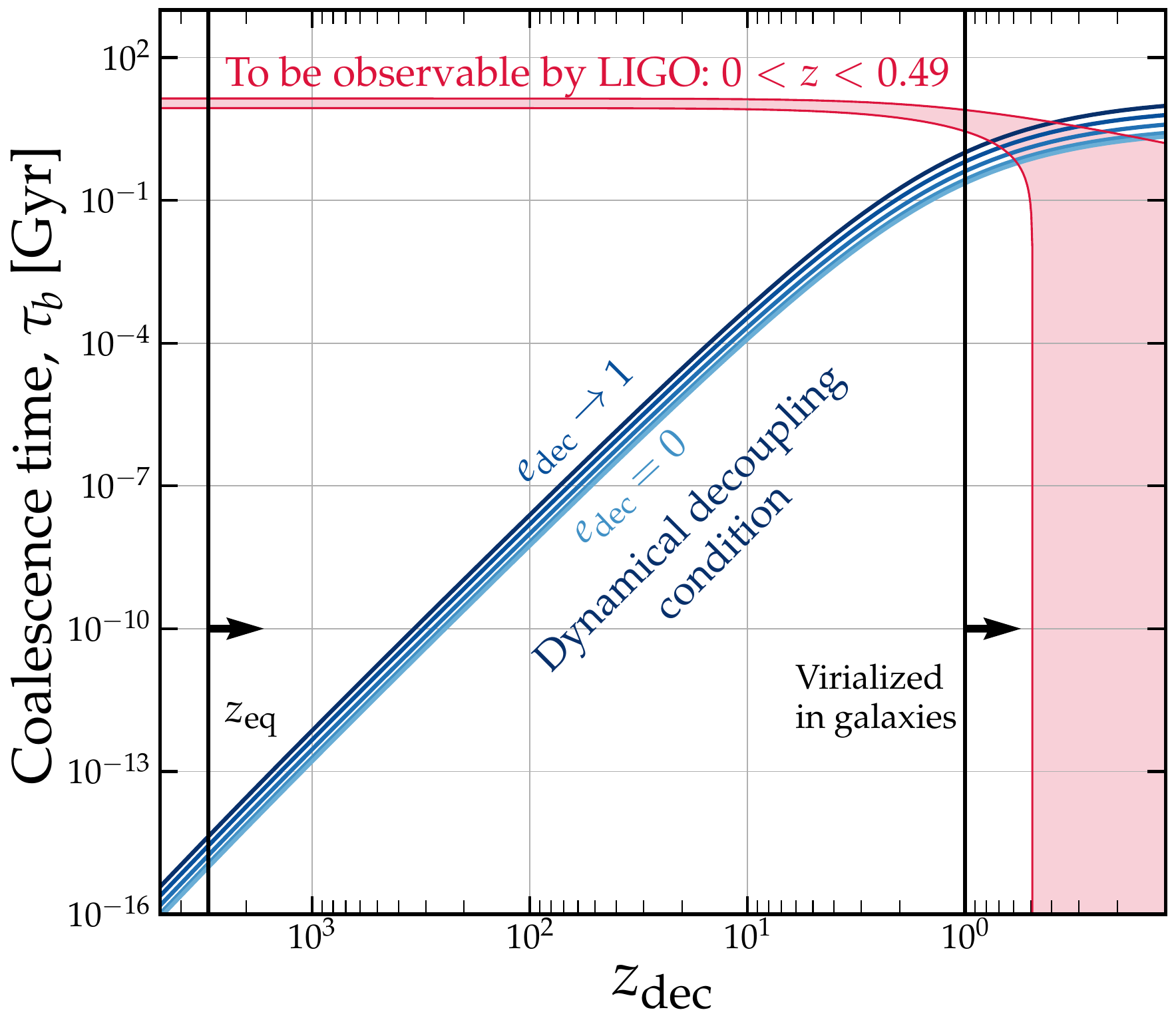}
    \caption{\label{fig:Coalescencetime} PBH binary coalescence times, as calculated in this work. For a given value of eccentricity, $e_{\rm dec}$ (equally spaced between 0 and 1), the coalescence time for binaries decoupling at redshift $z_{\rm dec}$ is given by the respective blue line. These coalescence times are what are \emph{required} for black holes embedded in expanding spacetime to emit enough energy in the form of gravitational waves to remain decoupled from the Hubble flow which tries to pull them apart.  In contrast, the coalescence times needed to be observed by LIGO/Virgo are enclosed by the red region. The binary needs to have merged between the farthest lookback time of the LIGO/Virgo (around $z\sim0.49$) and today. All binaries decoupling prior to galaxy formation will have merged well before they could be observed by LIGO/Virgo.}
\end{figure}

The final step is to apply the conditions written down previously to the population of PBH binaries we have just described. For the second ``dynamical'' condition eq.(\ref{dec2}) --- which is more restrictive than the static condition eq.(\ref{dec1}) --- we need to recalculate the binary energy loss following ref.~\cite{Peters:1964zz}, accounting for the quasi-local Misner-Sharp mass and the Hubble expansion rate. In the leading quadrupole approximation we can ignore some subleading terms proportional to the expansion rate and its time derivatives to find, 
\begin{align}
\dot E =-\frac{32}{5}\frac{G^4M^3\mu^2a^5}{\mathfrak{a}^5(1-e^2)^{7/2}}\left(1+\frac{73}{24}e^2+\frac{37}{96}e^4\right).
\end{align}
Hence, eq.(\ref{dec2}) implies for binaries decoupling prior to virialization:
\begin{align}\label{zrad}
\left(1+z_{\mathrm{dec}}\right)^3 H(z_{\mathrm{dec}}) < \frac{1}{\tau_b}\frac{96}{425}\left(1+\frac{73}{24}e_{\mathrm{dec}}^2+\frac{37}{96}e_{\mathrm{dec}}^4\right) \, .
\end{align}
The relationship between $\tau_{\rm b}$ and $z_{\rm dec}$ is shown in figure~\ref{fig:Coalescencetime}. Remarkably, there is no explicit dependence on the PBH mass; hence the assumption of a monochromatic mass function may not be that restrictive. 

Rearranging to evaluate $\tau_b$, we find that the dynamical decoupling condition implies that the PBH binaries must be so close together to decouple that their eventual coalescence times are very short. Taking the example of $e_{\mathrm{dec}}=1$ (which gives the longest $\tau_b$):
\begin{align}
    &\tau_{\rm b}<135~\mathrm{s~~~~~~}~~\left(z_{\mathrm{dec}}=z_{\rm eq}\approx 3000\right) 
    , \\
\intertext{for binaries that decoupled in the radiation era, and}
    &\tau_{\rm b}< 1.0~\textrm{Gyr~~~}~~\left(z_{\mathrm{dec}}=1\right) \, ,
\end{align}
for binaries that decoupled in the era of matter domination before PBHs are virialized in galactic halos. We thus conclude that the binaries that decouple in the radiation- and most of the matter-dominated eras merge well before the present day.

Figure~\ref{fig:Coalescencetime} shows the coalescence times as a function of $z_{\rm dec}$ and $e_{\rm dec}$. The reason why this result eliminates the LIGO/Virgo constraints on PBHs can be seen by comparing the gulf between the required $\tau_b$ (blue lines), and the observability window for LIGO mass black hole mergers (red shaded region). This window encloses the coalescence time that a binary decoupling at $z_{\rm dec}$ would need to have to count towards the mergers observed by LIGO/Virgo. This window is bounded from below by the requirement that the binary has not merged prior to the highest redshift merger (the very distant GW170729 at $z\sim 0.49$~\cite{LIGOScientific:2018mvr,Chatziioannou:2019dsz}), but also not so long that the binary still has not merged by the present day. The required coalescence times derived in this work fall short of this window by many orders of magnitude, right up until well past $z\sim1$. Beyond this point, galaxies will have formed, and the PBHs will be virialized in their halos. The merger rate in this environment is then controlled by an entirely different set of conditions~\cite{Bird:2016dcv}. This demonstrates that these early-universe PBH binaries would not be able to contribute to the merger rate or stochastic background~\cite{Mandic:2016lcn,Wang:2016ana} observed by LIGO/Virgo. Consequently, the bounds~\cite{Sasaki:2016jop, Ali-Haimoud:2017rtz} set on the PBH dark matter fraction based on this population of binaries can be \textit{completely} evaded, thus re-opening the possibility for PBHs in the LIGO-mass range. 

\section{Conclusion}\label{sec:conc}
By revisiting the PBH merger rate to account for a more adequate description of black holes in their surrounding spacetime, we have found that the merger rate constraints on the abundance of dark matter PBHs are evaded entirely. The reason for this is that PBH binaries which are able to form before galaxy formation must have had small initial separations and so would have merged well before the present time. The remaining PBHs --- which represent the vast majority of the population --- will not form binaries until galaxy formation under an alternative set of conditions.

To arrive at this conclusion we adopted the Thakurta metric as a more adequate description of PBHs in the early universe. This spacetime has a time-dependent quasi-local effective mass, so we must revise the conditions for a black hole binary to decouple from the Hubble flow. The two conditions, eqs.(\ref{dec1}) and (\ref{dec2}) --- labeled ``static'' and ``dynamical'' respectively--- are stronger than previously considered, implying a substantially lower binary merger rate in the window of cosmic time covered by LIGO/Virgo's observed mergers. This revitalizes the possibility that $\sim (30$--$100)~M_{\odot}$ mass PBHs could constitute all of the dark matter. The leftover PBHs that have not yet decoupled by $z\sim1$ will have been virialized in galactic halos. The merger rate for these PBHs will then follow the calculation presented in ref.~\cite{Bird:2016dcv}, which is intriguingly consistent with LIGO/VIRGO's observed merger rate.

Our conclusion represents a dramatic shift in the prospects for PBHs in cosmology. However, it is only a consequence of attempting to improve the description of black holes in the early Universe in a manner consistent with general relativity. Therefore our conclusion inspires a much more general message. Regardless of whether or not the Thakurta metric is truly the most adequate description of PBHs---and searches for better alternatives, if they exist, should certainly be encouraged---what we demonstrate here is that there are important and wide-ranging phenomenological consequences that stem solely from the choice of metric. These consequences must be taken seriously, both because they are potentially large, but more importantly because previous descriptions of PBHs have clear and unambiguous problems. So as well as advocating for the Thakurta metric specifically, we also identify that the problem of how to treat PBHs in expanding backgrounds is one that is worth more careful attention that it has been given previously. 

The implications of this work may extend beyond the LIGO/Virgo merger rate bounds, and could also invalidate numerous other constraints on PBH dark matter. Following essentially the same arguments, constraints on PBH dark matter from the stochastic gravitational-wave background~\cite{Mandic:2016lcn,Wang:2016ana} should similarly be invalidated. Additionally, the non-detection of subsolar-mass black hole mergers places strong constraints on the fraction of dark matter in subsolar-mass PBHs~\cite{Abbott:2018oah,Authors:2019qbw}. Since our final result is free of any mass dependence, it is safe to presume that these constraints will also be evaded.

The next step to decisively open up the possibility for dark matter in the form of LIGO-mass PBHs, is to address the remaining astrophysical bounds~\cite{Brandt:2016aco, Koushiappas:2017chw,Lu:2020bmd,Stegmann:2019wyz,Hektor:2018qqw,Poulin:2017bwe,Manshanden:2018tze,Gaggero:2016dpq, Inoue:2017csr}. Astrophysical uncertainties aside~\cite{Brandt:2016aco,Ali-Haimoud:2016mbv,Green:2017qoa,Bosch-Ramon:2020pcz}, many of these constraints will also be affected by a time-varying effective mass. In particular, since a black hole's accretion rate is proportional to its mass, early-universe constraints on PBH accretion could be weakened substantially. This may ultimately show that (30--100)~$M_\odot$ PBHs can constitute a large fraction, or possibly even the totality, of the dark matter in the Universe. We will address this issue in a follow-up study.
%So for example, the constraints set by the EDGES 21~cm absorption signal~\cite{Hektor:2018qqw} --- in which baryons at $z\approx 17$ would be heated by such accretion --- could be weakened. 
%There are also stronger constraints from CMB anisotropies, if one also assumes that the accretion geometry is disk-like instead of spherical~\cite{Poulin:2017bwe}. Efficient disk accretion requires that PBH binaries are formed early, so that enough angular momentum is supplied. Our results however imply that the fraction of decoupled PBH binaries is strongly suppressed at matter-radiation equality, so the assumption of accretion disk formation may not be justified. On top of this, the smaller active mass should also act to lower the overall accretion rate and alleviate the bounds further.
%A closer look at the effect of Misner-Sharp masses on accretion constraints is a natural next stage to fully open up the LIGO-mass window for PBH dark matter.

\section*{Acknowledgements}
We thank S\'ebastien Clesse for helpful comments on the manuscript and Lang Lui for spotting a sign typo. This article has been assigned the document number LIGO-P2000304. The work of AK was partially supported by Shota Rustaveli National Science Foundation of Georgia (SRNSFG) through the grant DI-18-335. The work of M.S. is supported in part by the Science and Technology Facility Council (STFC), United Kingdom, under the research grant ST/P000258/1.

\bibliography{bib.bib}
\bibliographystyle{bibi}

\end{document}